\documentclass[aps,onecolumn,nofootinbib,superscriptaddress]{revtex4}

\usepackage{calrsfs}
\usepackage{ulem}
\usepackage{graphics}
\usepackage{epstopdf}
\usepackage{bm}
\usepackage{bbm}
\usepackage{amssymb}
\usepackage{epstopdf}
\usepackage{graphicx}
\usepackage[caption=false]{subfig}
\usepackage{amsmath}
\usepackage{color}
\usepackage{array}

\begin{document}

\title{Snapshot hyperspectral imaging with quantum correlated photons}

\author{Yingwen \surname{Zhang}}
\affiliation{National Research Council of Canada, 100 Sussex Drive, Ottawa ON Canada, K1A0R6}

\author{Duncan \surname{England}}
\email{Duncan.England@nrc-cnrc.gc.ca}
\affiliation{National Research Council of Canada, 100 Sussex Drive, Ottawa ON Canada, K1A0R6}

\author{Benjamin \surname{Sussman}}
\affiliation{National Research Council of Canada, 100 Sussex Drive, Ottawa ON Canada, K1A0R6}
\affiliation{Department of Physics, University of Ottawa, Ottawa, Ontario, K1N 6N5, Canada}

\begin{abstract}
Hyperspectral imaging (HSI) has a wide range of applications from environmental monitoring to biotechnology. Current snapshot HSI techniques all require a trade-off between spatial and spectral resolution and are thus unable to achieve high resolutions in both simultaneously. Additionally, the techniques are resource inefficient with most of the photons lost through spectral filtering. Here, we demonstrate a snapshot HSI technique utilizing the strong spectro-temporal correlations inherent in entangled photons using a modified quantum ghost spectroscopy system, where the target is directly imaged with one photon and the spectral information gained through ghost spectroscopy from the partner photon. As only a few rows of pixels near the edge of the camera are used for the spectrometer, almost no spatial resolution is sacrificed for spectral. Also since no spectral filtering is required, all photons contribute to the HSI process making the technique much more resource efficient.
\end{abstract}
\maketitle

\section{Introduction}

Utilizing the properties of quantum entangled photons to enhance the performance of imaging techniques has been an active area of research in recent decades. Sub-shot noise imaging~\cite{Brida2010,Samantaray2017}, quantum ghost imaging~\cite{Pittman1995,Shapiro2012,Padgett2016}, imaging with undetected photons~\cite{Lemos2014}, and infra-red microscopy~\cite{Paterova2020} are just some examples of the imaging techniques made possible by the unique properties of entangled photons. With recent improvements in source and detector technologies, these techniques are becoming increasingly practical, and integration timescales have improved from hours to minutes, or even seconds~\cite{Zhang2022}. In this work, we showcase another advantage of quantum correlated imaging that may provide a viable alternative to classical methods. By applying the concepts of ghost spectroscopy~\cite{Scarcelli2003,Yabushita2004,Janassek2018,Ryczkowski2021} to quantum imaging, we show that snapshot hyperspectral imaging can be achieved by measuring the position of one photon from an entangled pair, and inferring its wavelength by making a spectral measurement on the other photon. With the use of a time-tagging camera, we demonstrate that a $\sim$20 kilopixel image with $\sim$2\,nm spectral resolution over an 85\,nm range can be achieved at the single photon level within a reasonable data acquisition time of a few minutes.

Hyperspectral imaging (HSI) is a class of imaging techniques used to obtain the spectral information for each spatial location on an image and has a wide range of applications ranging from environmental and agricultural monitoring, biotechnology, medical diagnostics to food analysis~\cite{Guolan2014,Bacon2004,Gowen2007,Orth2015,Slonecker2010,Delalieux2009,Zhu2020}. Traditional HSI techniques rely on scanning either the spatial or spectral degree of freedom~\cite{Green1998,Nahum2000}, and can provide high resolution in both position and spectrum, at the expense of long acquisition times due to the precision scanning required. Alternatively, so-called snapshot hyperspectral imaging (SHSI) techniques\cite{Hagen2013} can acquire spectral data in a single shot, by sacrificing spatial resolution. For example SHSI can be achieved using spectral filter arrays either directly on the camera sensor~\cite{Lapray2014,Wang2019,Monakhova2020}, or in conjunction with a microlens array~\cite{Shogenji2004,Mathews2008,Horstmeyer2009}. In these schemes, because an array of spectral pixels are required for each spatial pixel, a trade-off exists, and the spatial resolution is reduced in proportion to the desired spectral resolution. SHSI techniques without relying on spectral filter arrays have also been demonstrated, such as using the spatial and spectral correlations in speckle patterns~\cite{Sahoo2017}, but the trade-off between the spatial and spectral information still exists. Finally, it is worth noting that, in terms of the number of photons collected, all of these schemes are resource inefficient. The vast majority of photons arriving at conventional HSI apparatus will be lost; either blocked in the scanning apparatus, or removed by spectral filtering. In the weak illumination regime, this inefficiency will lead to significantly increased integration time.

In our SHSI demonstration, we show that the issues with resolution and efficiency can be overcome by using entangled photon pairs. We will term this technique quantum correlation hyperspectral imaging (QCHSI).  In QCHSI, we employ the temporal and spectral (anti)correlations in entangled photon pairs created through the process of spontaneous parametric down-conversion (SPDC). One photon, the \textit{signal}, is sent to illuminate the target, which is imaged by a time-tagging camera. Its entangled partner photon, the \textit{idler}, which had no interaction with the target, is sent onto a spectrometer built around the same camera. The photon pairs are then identified through a second order timing correlation measurement based on the photon detection times. Through this process, the spectral information at each pixel will automatically emerge as a result of the inherent spectral anti-correlation between the two photons. As the spectral information can be collected on just a few rows of pixels on the edge of the camera, almost no spatial resolution is sacrificed in exchange for the spectral information. Furthermore, in this scheme, every photon within the HSI apparatus can contribute to the image, and none are removed by scanning or filtering. Hence the efficiency of the measurement is not fundamentally limited by the scheme and is only limited by practical considerations such as camera and grating efficiency.

\section{Experimental setup}
\begin{figure*}[htbp]
	\centering \includegraphics[width=1\textwidth]{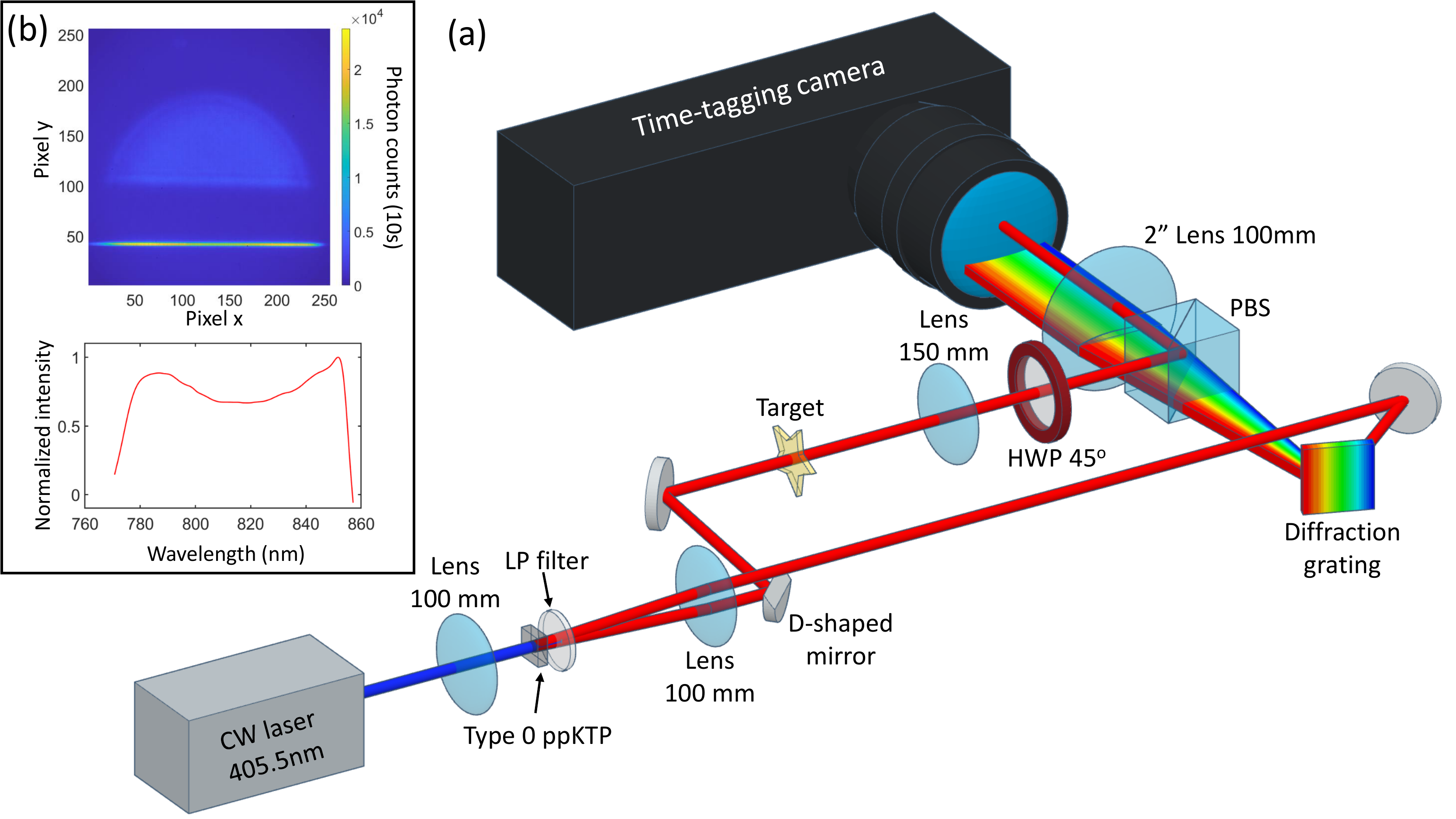}
	\caption{(a) A 405.5\,nm CW pump laser beam is focused onto a 1\,mm thick Type\,0 ppKTP crystal to generate temporal and spectrally (anti)correlated photon pairs. A long-pass~(LP) spectral filter is used to filter out the pump laser. The generated photon pairs are first collimated by a lens before split into two paths by a D-shaped mirror. The signal photon is sent to illuminate a target that is imaged onto the time-tagging camera. The idler photon is sent to a diffraction grating with the diffracted photons focused onto the camera by a lens, placed at one focal length from the grating and camera. The two photons are each imaged onto a different section of the camera with the aid of a polarizing beam-splitter (PBS) and a half-wave plate (HWP) rotated at $45^\circ$ placed in the path of the signal photon. (b) shows an raw image of the two beams captured on the camera accumulated over 10 seconds with no target present and the spectrum obtained from the cross-section of the spectrum arm.}
	\label{Fig1}
\end{figure*}
The experimental setup for QCHSI is shown in Fig.~\ref{Fig1}. A 405.5\,nm continuous wave (CW) laser beam is focused onto a 1\,mm thick Type~0 periodically poled potassium titanyl phosphate~(ppKTP) crystal to generate, through the process of SPDC, temporally correlated and spectrally anti-correlated photon pairs, namely the signal and idler photons. The SPDC photons have a spectral bandwidth of $\sim$100\,nm centered around 811\,nm. After filtering out the pump with a long-pass spectral filter, the SPDC beam is collimated by a 10\,cm focal length lens. The signal and idler photons are then separated using a D-shaped mirror (a 50:50 beam-splitter can also be used here to create a circular shaped beam but at the cost of losing 50$\%$ of all coincidence events). The signal photons are used to illuminate a target sample which is then imaged onto the time-tagging camera (TPX3CAM~\cite{Nomerotski2019,ASI} where the arrival time and position of each photon is tagged. The idler photon is sent onto a spectrometer built around a diffraction grating, lens, and the camera. 

The spectrometer is calibrated (see supplementary information) so that pixel number can be mapped to wavelength, therefore we can tag both the arrival time and wavelength of each idler photon. The two photon beams are recombined, but slightly displaced, by a a half-wave plate rotated at $45\deg$ in one of the paths and a polarizing beam-splitter placed just before the camera, such that each beam will be imaged onto a different location of the camera. A $809\pm41$\,nm spectral filter (Semrock FF01-809/81-25) is placed over the camera to block-off background light from outside the spectral range of interest. The TPX3CAM system has a spatial resolution of $256\times256$ pixels with a pixel pitch of 55\,$\mu$m and an effective temporal resolution of $\sim7$\,ns on each pixel. We utilize this feature to time-tag every photon detected on each pixel and use this information to identify temporally correlated photon pairs. After identifying the coincident photons, the per-pixel spectral information of the sample will automatically emerge as a result of the inherent spectral anti-correlation between the signal and idler photons.

To recover the spectral response of the target, some corrections to the measured coincidence spectrum need to be made. The first is to remove accidental coincidence events due to random detection overlap between uncorrelated photons. For accidental coincidences between pixel $n$ of the idler beam and pixel $m$ of the signal beam, this is given by
\begin{equation}
    C_{n,m}^{\text{acc}} = N_n N_m\tau,
\end{equation}
where $N_{n(m)}$ is the number of photons detected per second in pixel $n(m)$ and $\tau$ is the coincidence gating time, which for this experiment is set at 20\,ns. Thus, the total number of accidental coincidences in a pixel $m$ is given by
\begin{equation}
    C_{m}^{\text{acc}} = \sum_n N_n N_m\tau.
\end{equation}

The second is that since the measured spectrum is of the idler photon and the two photon spectrum is anti-correlated, the coincidence spectrum obtained is in fact inverted from what is seen by the signal photon. This inversion is corrected using the energy conservation relation
\begin{equation}
    \frac{1}{\lambda_p} =  \frac{1}{\lambda_s} + \frac{1}{\lambda_i},   
\end{equation}
where the $\lambda_p$, $\lambda_s$ and $\lambda_i$ are the wavelengths of the pump, signal and idler photons. Also due to this wavelength inversion, one should not send both photons to the target and split them afterwards in the hopes of getting a better signal to noise ratio. The coincidence spectrum obtained this way will be an overlapped spectrum of the target's spectral response and its inverse spectral response. The spectrum is calibrated by matching the measured spectral shape to the fall-off edge of various spectral filters. 

Finally, we note that, in common with all spectrometers, the spectrum determined after these corrections are convolutions of the target spectrum with the instrument response (mainly from the camera and grating), to correct for this will require a careful calibration of the detection system, which is beyond the scope the this work. Details on this effect are outlined in the supplementary information.


\section{Results}
\begin{figure*}[htbp]
	\centering \includegraphics[width=1\textwidth]{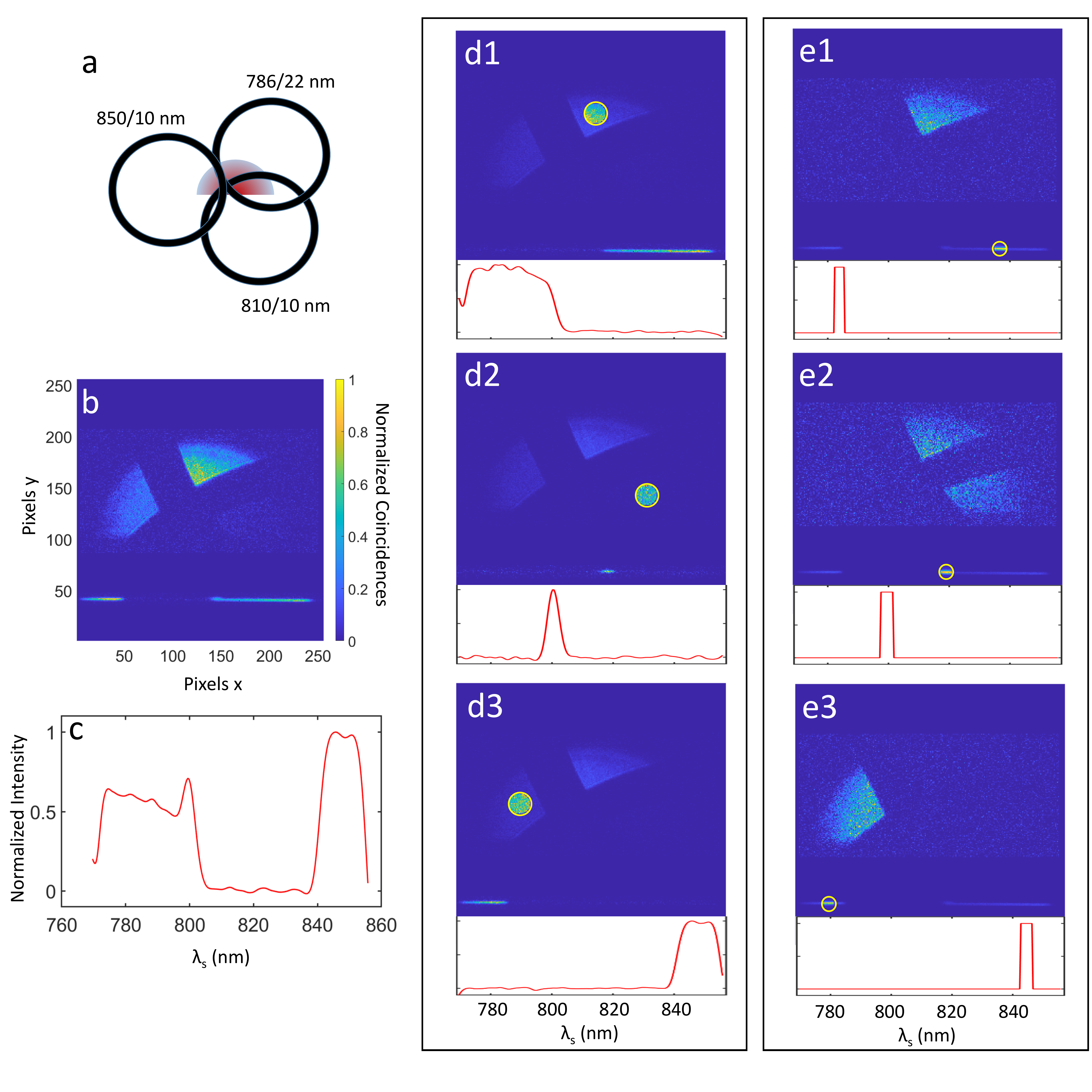}
	\caption{a - Orientation of the 3 band-pass filters ($786\pm11$\,nm, $810\pm5$\,nm, $850\pm5$\,nm) with respect to the imaging beam. b - Image of the twin beams after temporal correlation analysis.  c - Average signal photon spectrum as calculated by energy conservation. d1-d3 - Correlated spectrum of selected regions (highlighted by the yellow circle) in the imaging signal arm with the corresponding signal photon spectrum shown in the insets. As reference, a faint background of the full beam is shown in the images. e1-e3  Correlated image of selected regions (highlighted by the yellow circle) in the spectrometer arm with the insets showing the corresponding wavelength region selected for the signal photon. A faint background of the full correlated spectrum is shown in the images as reference. Data acquisition time is 100\,s and all displayed images and spectrum are background subtracted. As the maximum intensity of the idler spectrum is much brighter than the signal beam, the two are normalized separately in all the displayed images in order to make features in the signal beam more visible. The displayed signal spectrum have been smoothed through a cubic spline.}
	\label{Fig2}
\end{figure*}

As a proof of principle demonstration of QCHSI, we placed three different spectral band-pass filters - $786\pm11$\,nm (Semrock FF01-784/22-25), $810\pm5$\,nm (Semrock FF01-810/10-25) and $850\pm5$\,nm (Semrock FF01-850/10-25) - in the path of the signal photon, their orientation with respect to the beam is shown in Fig.~\ref{Fig2}(a). The image obtained after coincidence measurement is performed between the full signal beam and the idler spectrum is shown in Fig.~\ref{Fig2}(b) with the corresponding signal photon spectrum shown in Fig.~\ref{Fig2}(c). 

In Fig.~\ref{Fig2}(d), we show the spectrum at various regions of the target. This is obtained when coincidence events between the highlighted region of interest (ROI) on the signal beam and the full idler beam spectrum is considered. The reverse is shown in Fig.~\ref{Fig2}(e), where coincidence events between a small ROI in the idler spectrum and the full signal beam is considered. Here, only regions on the target with signal photons spectrally correlated with the idler photons in the ROI will be lit up.

\begin{figure*}[htbp]
	\centering \includegraphics[width=1\textwidth]{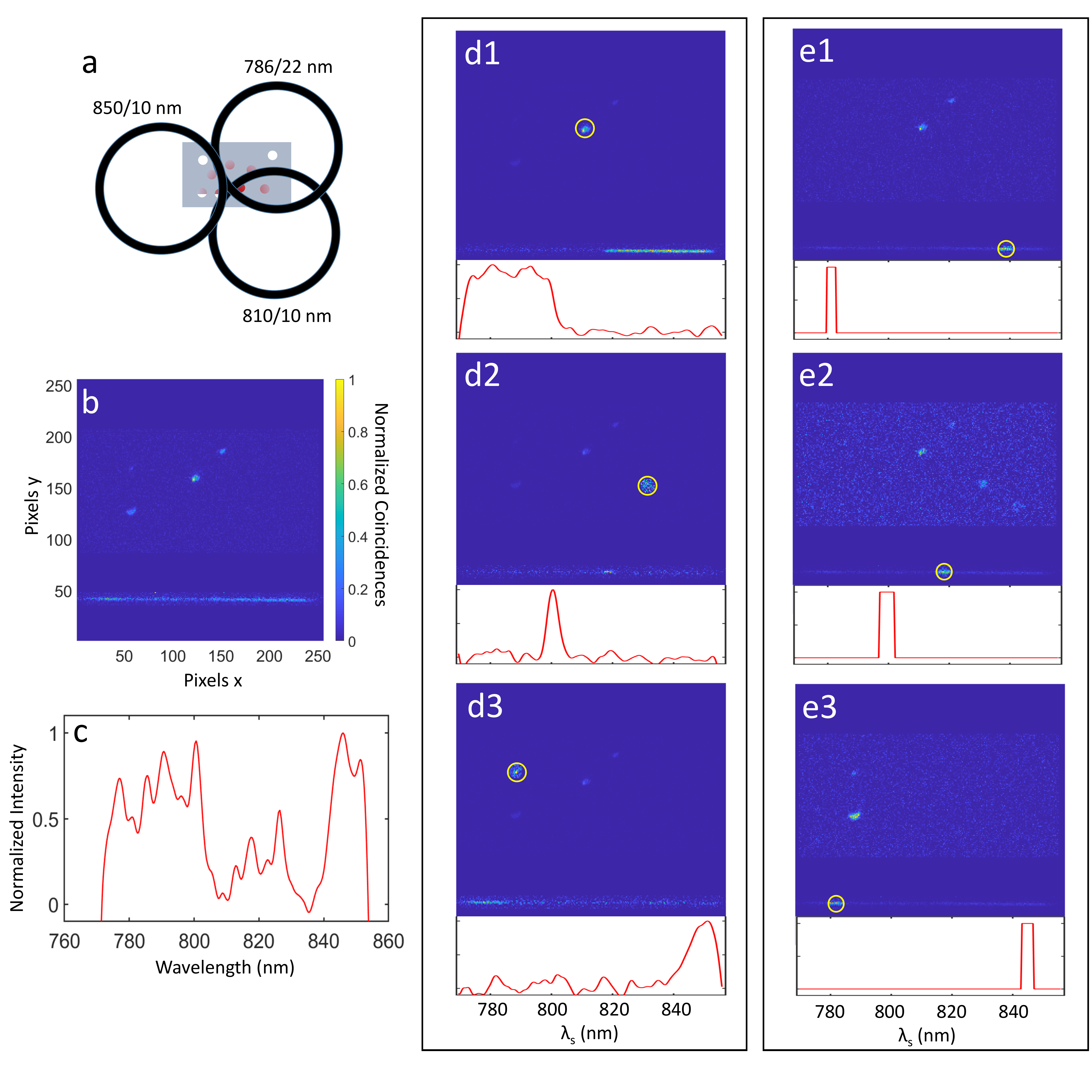}
	\caption{a - Orientation of the 3 band-pass filters  ($786\pm11$\,nm, $810\pm5$\,nm, $850\pm5$\,nm) with respect to the imaging beam. Over the beam is a piece of paper with small punctured holes of $\sim$0.2\,mm in diameter. b - Image of the twin beams after time correlation analysis. c - Average signal photon spectrum as calculated by energy conservation. d1-d3 - Correlated spectrum of selected regions (highlighted by the yellow circle) in the imaging signal arm with the signal spectrum shown in the insets. As reference, a faint background of the full beam is shown in the images. e1-e3  Correlated image of selected regions (highlighted by the yellow circle) in the spectrometer arm  with the insets showing the corresponding wavelength region selected for the signal photon. A faint background of the full correlated spectrum is shown in the images as reference. Data acquisition time is 270\,s and all displayed images and spectrum are background subtracted. As the maximum intensity of the idler spectrum is much brighter than the signal beam, the two are normalized separately in all the displayed images in order to make features in the signal beam more visible. The displayed signal spectrum have been smoothed through a cubic spline.}
	\label{Fig3}
\end{figure*}

In Fig.~\ref{Fig3}, a piece of paper with small punctured holes of $\sim$0.2\,mm in diameter is placed in front of the spectral filters to simulate a situation where the signal photons are reflected by or transmitted through very small targets (relative to the beam size). We see in Fig.~\ref{Fig2}(b) and (c) that due to the high noise from the small amount of transmitted signal photons, only a few bright spots can be seen on the target and features in the spectrum are also mostly washed out by noise. However, as seen in Fig.~\ref{Fig2}(d), if coincidences from just a few bright spots are considered, the spectrum can still be recovered with good visibility. On the other hand, as seen in Fig.~\ref{Fig2}(e), if only coincidences from a small ROI in the idler spectrum is considered, all points in the target with correlated wavelength is lit up, some of which are too faint to be visible in Fig.~\ref{Fig2}(b).

\section{Discussion}
The data acquisition time of the current system is in the order of a few minutes, this is a direct result of the limited capability of the camera system which has a quantum efficiency of $\sim$5$\%$, a temporal resolution of $\sim$7\,ns, and a saturation limit of $\sim$10$^7$ photons per second. With expected advances in camera technology, each parameter could be improved by a factor of 10 in the near future; this would be enough to allow real-time spectral imaging using QCHSI.

The spectral resolution of the QCHSI technique is determined by the bandwidth of the pump laser and the resolution of the spectrometer. In this case, the pump is a narrow linewidth continuous wave laser so the resolution limit is dominated by the spectrometer setup. Based on the thickness of the spectral band and the width of the spectral peak in the overlap region between the $786\pm11$\,nm and $810\pm5$\,nm filters, we estimate the spectral resolution of our system to be $\sim$2\,nm. With improvements to the pair source and the imaging optics, we expect the resolution of this setup could approach $0.7$\,nm as measured in a previous experiment~\cite{Zhang2021}. 

In this initial demonstration, type-0 degenerate SPDC was used such that each photon mode is centred at 810\,nm, with a spectral bandwidth of $\sim$100\,nm. The bandwidth could be further increased by reducing the crystal thickness for larger spectral coverage or, for increased spectral resolution, narrower bandwidth photons could be generated by type-II SPDC. Furthermore, non-degenerate pair sources could be used to interrogate samples in very different spectral windows~\cite{Aspden2015}. In this demonstration, the same camera is used for both the image and the spectrum for simplicity. But, more generally, the spectral information could be obtained using a second camera, or a separate device such as a 1D avalanche photo-diode array~\cite{Johnsen2014}, or a time-of-flight fiber spectrometer~\cite{Avenhaus2009}. Finally, we note that QCHSI could also be operated in ghost imaging mode, where the target is placed in the spectrometer arm instead of the imaging arm, however it can be difficult to accurately find the image plane of the ``ghost" target in the imaging arm using this approach.

Though we have demonstrated significant advantages of using QCHSI, there are still inherent limitations in the technique which cannot be solved by improved detectors. Principally, QCHSI is an active imaging technique where the target must be illuminated with quantum correlated photons generated by the observer, it cannot work passively using background light. Most conventional SHSI techniques can work both with active and passive illumination. Secondly,  QCHSI it is not suitable for imaging fluorescent targets because the wavelength change imparted by fluoresence in the signal arm  will not be detected by observing the idler arm.

Our demonstration can be potentially be performed using classical ghost spectroscopy which has been demonstrated in recent years, using either the spectral correlation present in specialized thermal sources~\cite{Janassek2018} or through computational methods using a programmable spectral filter~\cite{Ryczkowski2021}. Due to the nature of classical ghost spectroscopy needing to scan hundreds of spectral patterns, we expect the data acquisition time will be similar to that presented in this work.

\section{Conclusion}
In conclusion, we have demonstrated a proof of concept SHSI system utilizing the strong temporal and spectral (anti)correlations inherent in entangled photon pairs generated through the process of SPDC. With this system, almost no spatial resolution will need to be sacrificed in exchange for the spectral as in most conventional SHSI techniques, permitting for the maximum spatial and spectral resolution allowed by the detection camera and imaging system to be achieved simultaneously. Utilizing the background noise reduction provided by the temporal correlation measurement and the higher photon efficiency of the technique (no photon losses through scanning or spectral filtering) makes this technique well suited for applications requiring weak illumination such as covert illumination or imaging of light sensitive samples. Due to limitations in detection technology, the performance of QCHSI is currently still far from matching commercial SHSI cameras in terms of spatial resolution and data acquisition time, but we believe this issue can be overcome in the near future with the advent of improved single photon detection technologies~\cite{Timepix4,Nomerotski2020,Morimoto2020}.

\section*{Supplementary Information}
\begin{figure*}[htbp]
	\centering \includegraphics[width=1\textwidth]{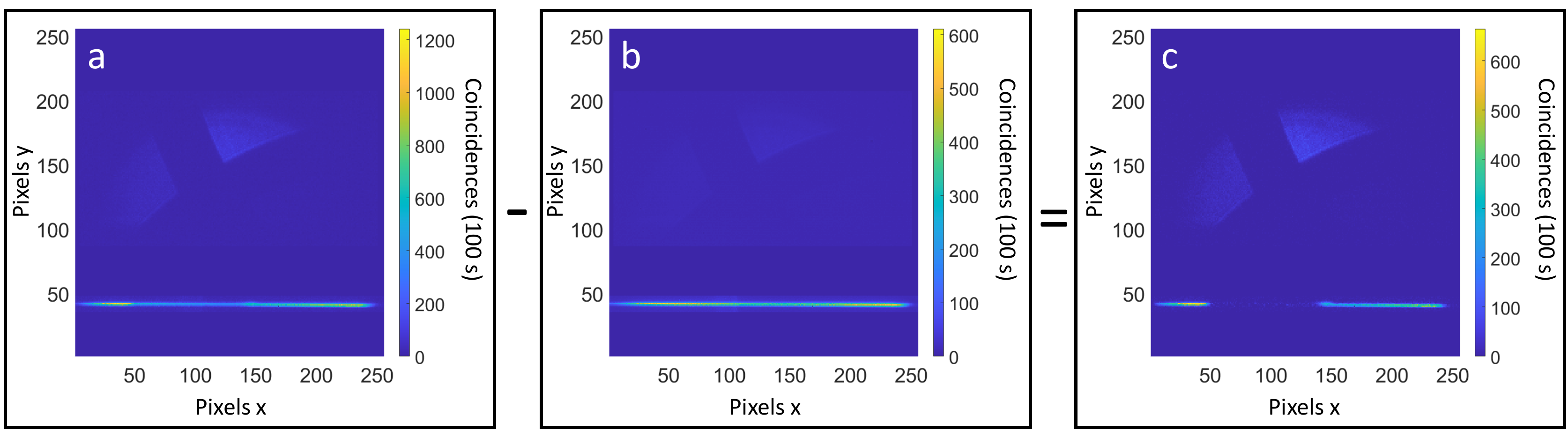}
	\caption{Background subtracted coincidence image~(c) is obtained by subtracting the background image~(b), determined from eq.~(2) of the main text, from the raw coincidence image~((a).}
	\label{SFig1}
\end{figure*}
Background subtraction in the coincidence image is demonstrated in Fig.~\ref{SFig1} where the background image is determined pixel wise according to eq.~(2) of the main text.

\begin{figure*}[htbp]
	\centering \includegraphics[width=1\textwidth]{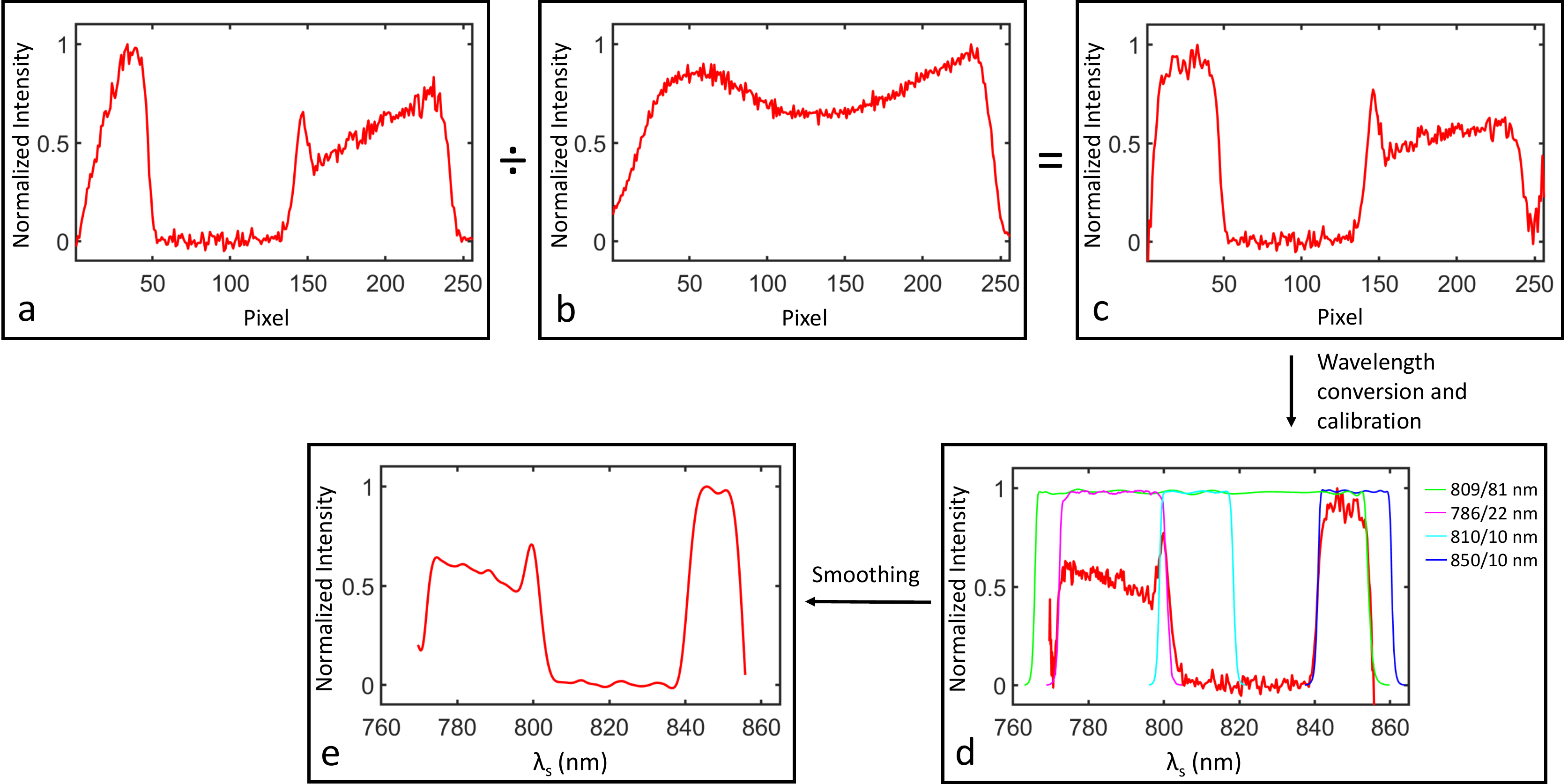}
	\caption{The directly measured idler photon spectrum~(b) is divided by the raw coincidence idler photon spectrum~(a) to obtain a corrected coincidence spectrum of the idler photon~(c). The pixel number is then converted to wavelength and inverted according to energy conservation, given by eq.~(3) of the main text, to obtain the spectrum of the signal photon. The pixel to wavelength conversion is calibrated according to the edges of multiple spectral filters as seen in~(d). The final spectrum~(e) is obtained by smoothing (d) with a cubic spline fit in order to remove the large fluctuations seen on the edges of the spectrum, a result from division by small numbers during the correction from (a) to (c). }
	\label{SFig2}
\end{figure*}

The steps to obtain the final signal photon spectrum, as discussed at the end of the ``Experimental setup" section of the main text, is illustrated in Fig.~\ref{SFig2}.

To calculate the absorption spectrum of the target $S_{\text{targ}}(\lambda_s)$, we must account for the spectral response of the optics, and camera, and for the spectral profile of the SPDC photons. The measured coincidence spectrum $S_{\text{coinc}}(\lambda_i,\lambda_s)$ in Fig.~\ref{SFig2}(a) is related to the spectral profile of each element as follows
\begin{equation}
    S_{\text{coinc}}(\lambda_i,\lambda_s) = S_{\text{SPDC}}(\lambda_i,\lambda_s) \left[\eta_{\text{cam}}(\lambda_i)\eta_{\text{grat}}(\lambda_i)\right]\left[\eta_{\text{cam}}(\lambda_s)S_{\text{targ}}(\lambda_s)\right],
\end{equation}
where $\eta_{\text{cam}}$ and $\eta_{\text{grat}}$ are the wavelength-dependent quantum efficiency of the camera, and grating efficiency respectively. We have ignored the spectral profile of simple optical elements such as lenses and mirrors which are mostly flat within the spectral region of this experiment. 
$\lambda_s$ and $\lambda_i$ are related through the conservation of energy as given by eq.(3) of the main text, i.e. $\frac{1}{\lambda_p} =  \frac{1}{\lambda_s} + \frac{1}{\lambda_i}$. The measured idler photon spectrum as seen in Fig.~\ref{SFig2}(b) is given by the product of the SPDC spectrum with the camera and grating efficiencies:
\begin{equation}
    S(\lambda_i) = S_{\text{SPDC}}(\lambda_i,\lambda_s) \left[\eta_{\text{cam}}(\lambda_i)\eta_{\text{grat}}(\lambda_i)\right].
\end{equation}
By dividing the coincidence spectrum by the idler spectrum as shown in Fig.~\ref{SFig2}, this returns an output spectrum Fig.~\ref{SFig2}(c) of the form:
\begin{equation}
    S_{\text{out}}(\lambda_i) = \frac{\eta_{\text{cam}}(\lambda_s)S_{\text{targ}}(\lambda_s)}{\eta_{\text{cam}}(\lambda_i)\eta_{\text{grat}}(\lambda_i)}.
\end{equation}
So we can see that this technique does not directly measure the spectrum of the target $S_{\text{targ}}$ but rather, in common with all spectrometers, measures the product of the target spectrum and the instrument response. With careful intensity calibration of the camera system and measurements of the grating efficiency, $S_{\text{targ}}$ could be determined, but this is beyond the scope of this work.

Conversion from position to wavelength in a spectrometer is slightly nonlinear, thus after inversion of the idler photon spectrum to obtain the spectrum of signal photon, a third order polynomial fit, given by $\lambda = ax^3+bx^2+cx+d$ (with $\lambda$ the wavelength and $x$ the pixel number), is made in order to match the obtained spectrum to the edges of the various spectral filters (with spectral response obtained from Semrock), this is seen in Fig.~\ref{SFig2}(d). The obtained fitting parameters are $a = -4.83\times10^{-6}$\,nm, $b = 1.78\times10^{-3}$\,nm, $c = 0.199$\,nm, $d = 770.5$\,nm.

\section*{Acknowledgements}
The authors are grateful to Antony Orth, Andrew Ridsdale, Denis Guay, and Doug Moffatt for stimulating discussions and technical support. This work was partly supported by Defence Research and Development Canada.

\bibliography{HyperspectralRef}

\end{document}